\documentclass[doublecol]{epl2} 
\title{Many-Body Localization Transition in Random Quantum Spin Chains  with Long-Range Interactions}
\shorttitle{Many Body Localization Transition in Random Quantum Spin Chains }
\author{{ N. Moure \inst{1} \and} S. Haas \inst{1,2} \and S. Kettemann \inst{2,3}}
\institute{ \inst{1}Department of Physics and Astronomy  
University of Southern California,
Los Angeles, CA 90089-0484 \\
 \inst{2} Department of Physics and Earth Sciences, Jacobs University
  Bremen, Bremen 28759, Germany \\
  \inst{3}
  Division of Advanced Materials Science, Pohang University
  of Science and Technology (POSTECH), Pohang 790-784, South Korea} 

\abstract{
   While there are well established methods to study delocalization transitions of 
    single particles in random systems, it remains a challenging problem how to characterize
  many body delocalization transitions. 
 Here, we use 
a generalized  real-space renormalization group technique to study the anisotropic Heisenberg model
with long-range interactions, decaying with a power $\alpha$, which are generated by placing spins at random 
positions along the chain. This method permits a large-scale finite-size scaling analysis.
  We {examine} the full  distribution function  of the  excitation energy gap from the ground state
  { and observe a crossover with decreasing $\alpha$}.
 {At $\alpha_c$ }  the full   distribution coincides with a critical function.
    Thereby, we find   strong evidence for the existence of a many body localization transition in disordered  antiferromagnetic spin chains with long range  interactions.
}

\begin{document}
\maketitle




%
{\bf Introduction. --} Long-range interactions 
between local quantum degrees of freedom, 
 such as spins and quantum rotors, are ubiquitous
  in real materials, such as  doped semiconductors and glassy systems. 
Anomalous  magnetic properties of  doped 
 semiconductors,  e.g.  the  low-temperature power-law divergence of  their magnetic susceptibility, 
 are thought to arise  from  
    local magnetic moments,  positioned randomly  
  \cite{pwanderson0,pwanderson,mott,loehneysen,finkelsteinesr,senthil}. These moments are coupled
   in the insulating phase
 within a finite range, 
      limited by the localization length.
       In the metallic phase the  coupling becomes long-ranged, decaying 
        with a power of the distance between magnetic moments\cite{rkky}. 
        Low temperature 
 properties of a wide range of 
  glassy systems can 
  be modeled      by   2-level systems describing the excitations of  ions   tunneling
    between local potential minima\cite{reviewtls,Yu,levitov,prl1999}.
  Dipole-dipole interactions between their dipole moments 
  and  elastic coupling between them 
  lead  to  an effective model of  random long-range coupled Heisenberg spins.
   Recently,   there have   been experimental  
  indications of a novel quantum phase transition
   to a collective state      in such  a system\cite{prl1999}. 
  Thus,  a systematic analysis of 
   long-range 
   coupled quantum models is called for.
   
  {  The random spin-1/2 Heisenberg chain is considered to be the paradigm 
  of disordered systems whose low energy universal behavior is controlled by an infinite randomness fixed point where all spins are bound to randomly located singlets, 
 if the interactions are  antiferromagnetic and  short ranged\cite{dasguptama,bhattlee,fisher,fisher2}. 
 In  real materials the interaction between local magnetic moments
   is longer ranged. Thus,  it is an  important open question of practical importance, if  the strong disorder fixed point  becomes destabilized  
  and    a delocalization transition to extended spin excitations is induced with increasing interaction range. }
   
  Dynamics and relaxation in disordered systems  is characteristically different from ordered systems, as it involves distributions of relaxation times and  activation energies.   As noted  early on, the physics of random systems is   fully described  by probability distributions
   of quantities like the activation energy \cite{pwanderson}. An analysis based only on averages  is  likely to miss
   relevant physical processes such as rare events\cite{bhattfisher}. 
   In this article we
   implement  the real-space renormalization group method
   to investigate  random  quantum spin models  with long-range couplings by 
     analyzing  full  distribution functions of their  excitation energy gaps from the ground state, $\epsilon_1$.
When  states  are localized at different positions in space,   they are  uncorrelated
  and  one expects
  the  spacing  between neighbored energy levels, such as $\epsilon_1,$  to
   follow the  Poisson {level spacing } distribution.  In contrast,  extended states  overlap, causing
  power law  level repulsion, which results   in  the Wigner surmise distribution function of    $\epsilon_1$. { Therefore, the position and critical properties of delocalization transitions can be 
   characterized by   analyzing the distributions of  level spacings in their vicinity\cite{levelstatistics}. }
  
   The  non-interacting Anderson model of disordered fermions 
  with  long-range hoppings, decaying  with distance as
      $R^{-\alpha}$ is well known to    show an insulator-metal transition 
       as  function of the decay exponent $\alpha$.
   When $\alpha > d$,  ($d$ is the dimension), 
      all states are  localized\cite{fyodorov1996}.
       Localization means in this case that 
        for length scales $r> \xi$,  ($\xi$  the localization length), 
         the eigenfunctions decay  as $\psi(r) \sim r^{-\alpha}.$
   For $\alpha < d$ the eigenstates are extended,  $\psi(r)$ does not decay 
    with distance. 
    One can detect the transition by calculating the inverse participation ratio
     $I_2 = \int d{\bf r} |\psi({\bf r})|^4\sim L^{-\tau_2}$\cite{efetov,mackinnon,rmpmirlinevers}.
      For $\alpha < \alpha^0_c  =d$, one finds $\tau_2=d$ corresponding to extended
       states, whereas for $\alpha > d$, $\tau_2 =0$, corresponding to 
        localized states\cite{fyodorov1996}.
       When $\alpha =d$ 
        the system is  critical and  eigenfunction intensities 
         exhibit  
        multifractality, $\langle |\psi|^{2q} \rangle \sim L^{- d_q (q-1)-d}$,
         where $d_q$ is the multifractal dimension of the q-th moment\cite{multifractal,rmpmirlinevers}. 
        The critical  inverse participation ratio scales with $L$ with  power $\tau_2=3 d- 2 \alpha_0,$
         where $\alpha_0>d$ is the multifractality parameter, which depends on  system dimension and symmetries.   We note that there are other classes of  random models
        with long range interaction,
        which do show a delocalization transition at $\alpha < d$. 
         One such system is the model of non-interacting  fermions with  non-random long-range coupling and diagonal disorder\cite{nonrandomlr}. 
         Another example are the hierarchical models studied in Refs. 
         \cite{hierarchical}.
        Random banded matrices with  critical long-range coupling 
         have  been studied for $d=1$ as 
          paradigmatic models of  Anderson metal-insulator transitions (MIT), 
      allowing large length,  numerical finite-size scaling.

   {\bf Random Quantum Spin Chains. --}  Here, we  study random  quantum spin chains\cite{monthus} with long-range couplings $J_{ij} = J|{\bf r}_i-{\bf r}_j|^{-\alpha}$, where $ 0 < \alpha < \infty$. 
      It is expected that there occurs a many-body transition between 
       localized and delocalized states, at a critical $\alpha_c$.  
       However,   it is not yet known if $\alpha_c$  is  
       equal to  the non-interacting value $\alpha^0_c =d=1$.
     It has been shown 
     rigorously for clean spin chains with long-range exchange couplings  that  the ground state  has long-range order
         when   $\alpha < \alpha^*=2 d =2$, based on  an extension 
           of the Mermin-Wagner theorem\cite{bruno}. 
            Since disorder tends to suppress long-range order, 
     $ \alpha^*$ is expected to decrease with disorder towards
      smaller values or even to vanish.  
          Thus,  one can expect that 
           delocalization   occurs first at an upper critical  $\alpha_c$, before
            the  transition to an ordered
            state happens 
            at smaller $\alpha$, allowing for an intermediate phase, such as a spin glass phase.  
           The aim of this article is to identify and characterize the delocalization transition. 
             
   We consider the Hamiltonian of the random XXZ-Heisenberg Model
       \begin{equation}\label{spinchain}
       H= \sum_{i,j} \left[ J_{x i,j} ( {\bf S}_{x i} {\bf S}_{x j} +   {\bf S}_{y i} {\bf S}_{y j}  )  
          + J_{z i,j} {\bf S}_{z i} {\bf S}_{z j}    \right],
       \end{equation}
       { with $r_i, r_j$ representing } the N sites, where a spin is placed, as chosen randomly from a lattice of L sites at fixed density $N/L=0.1$ 
        with periodic boundary conditions. We assume antiferromagnetic coupling 
          between all pairs of sites $i,j$ with $J_{ij} = J|{\bf r}_i-{\bf r}_j|^{-\alpha}$. 
 Thus,  the  couplings $J_{ij}$ are randomly distributed with the typical coupling between nearest neighbor spins
  $J_{nn}=J (L/N)^{-\alpha}.$ The  coupling between any spins  in the chain cannot become smaller than the minimal coupling
  $J_{min}=J (L/2)^{-\alpha}.$
    
      {\it Jordan Wigner Transformation.}
    It is insightful to use     the Jordan-Wigner transformation which maps    the spin chain  Eq. (\ref{spinchain}) onto
    the Hamiltonian of  interacting fermions. 
   For  $J_z=0$ one thereby finds 
    \begin{equation}\label{jw}
       H= \sum_{i,j}  J_{x i,j} \left( c_i^+ c_j e^{i \pi \hat{n}_{ij}} + c_j^+ c_i e^{-i \pi \hat{n}_{ij}}  \right),
       \end{equation}
       where the operator $\hat{n}_{ij} =\sum_{i<n<j} c_n^+ c_n $ counts how  many fermions 
       are encountered while hopping between the sites $i$ and $j$. 
      For nearest neighbor hopping this is exactly the 
       Hamiltonian of  noninteracting  fermions with random hopping,
        which is known to  show the Dyson anomaly:
        the eigenfunctions in the center of the band
         decay spatially with a stretched exponential,
           $\psi(x) \sim \exp (- \sqrt{x/l_0}]), $ where $l_0$ is a small length scale\cite{dyson}. 
           Away from the band center the eigenfunctions decay exponentially with localization length $\xi$, which diverges at the band center as 
            $\xi \sim - \ln |\epsilon|.$
          The density of states is singular at half filling, 
           $\rho(\epsilon) = |\epsilon|^{-1} \ln |\epsilon|^{-3}$\cite{wegner}.
        Therefore,    the expectation value of the nearest level spacing shifts to 
         the small
             value $
    \Omega_N \sim \exp(-\sqrt{N}).$            
       For longer  range hopping 
    the interaction between the fermions 
     manifests itself 
      through the  fluctuating phase factors in the hopping amplitudes,
       making this a  challenging many body problem. 
 
      {\bf  Real Space Renormalization Group. --}   Let us therefore return to Eq.   (\ref{spinchain}) 
       and  apply the real-space renormalization group (RSRG)\cite{dasguptama,bhattlee,fisher,fisher2,monthus,refael,damle}
             procedure, which  enables us to  study larger systems
             numerically 
               than with exact  diagonalization\cite{haas1}.
             One starts with 
           the strongest coupled pair,  say  $(i,j)$, 
            which in its ground state 
             forms a singlet. 
              Taking the expectation value of   Eq.   (\ref{spinchain})
          in that singlet state, performing second-order perturbation 
             theory in the coupling { with  other  spins}\cite{sigrist}, one obtains an effective Hamiltonian where the coupling 
              between  spins $(l,m)$ is renormalized.  For the coupling 
               between  x (or y)- and   z-components one gets
              \begin{eqnarray} \label{jeff}
               (J^{x}_{lm})' &= &  J_{lm}^x - (J^x_{li}-J^x_{lj})(J^x_{im}-J^x_{jm})/(J^x_{ij}+J^z_{ij}),
               \nonumber \\
               (J_{lm}^{z})' &= &  J^z_{lm} - ((J^z_{li}-J^z_{lj})(J^z_{im}-J^z_{jm})/2 J^x_{ij}.
              \end{eqnarray}
                These renormalization rules are anisotropic and 
                 are valid as long as $J_x\gg 0$. 
                  For $J_x \rightarrow 0$, the Ising limit, 
                  the ground state becomes degenerate between the two N\'eel states, 
                   and degenerate perturbation theory yields  the 
                   correct  RG rules \cite{isingrgrules}. 
                    When initially all couplings have the same anisotropy $\gamma = J_{ij}^{z}/J_{ij}^{x}$, 
                     the renormalization rules Eqs. (\ref{jeff}) changes the anisotropy of
                      the renormalized pair $lm$  to 
                     \begin{equation}
                     \gamma_{lm} =   \gamma^2 \frac{1+\gamma}{2}.
                     \end{equation}
                      Thus, while the isotropic chain $\gamma =1$ is a fixed point, 
                       any anisotropy drives the chain to 
                       i) the XX random singlet phase  $\gamma \rightarrow 0$
                        for $\gamma <1$, ii) the Ising antiferromagnet
                         with staggered magnetization in z-direction,   $\gamma \rightarrow \infty$
                        for $\gamma >1$\cite{fisher2}. 
                    We will focus here on the regime $ \gamma \rightarrow 0$,  $J_z \rightarrow 0,$ the XX-limit with  long-range interactions in order to  see
                     if the random singlet phase is stable 
                      when longer range interactions are added and to  search for a possible  many body localization transition{\cite{basko,znidaric,garel,bardarson}}.

                        \begin{figure}[t]
                     \vspace{-.5cm}
{\includegraphics[width=0.4 \textwidth]{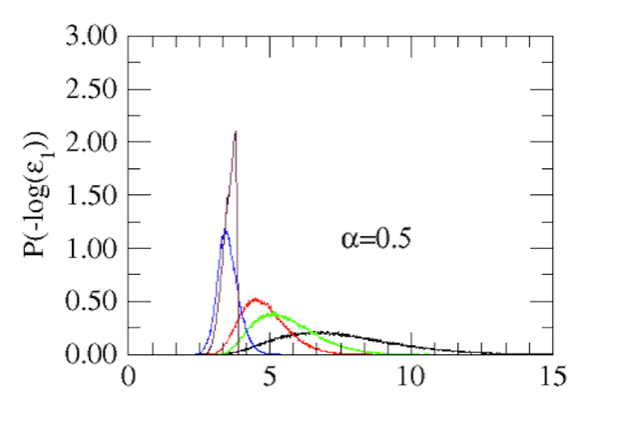} }
\vspace{-.5cm}
\caption{ Distributions
 of excitation energy $\epsilon_1$ 
  in  random XX-spin chains for $\alpha=0.5$  ($N=128$).
True long-range coupling (brown curve)  compared with 
nearest-neighbor  (black), 
next-nearest-neighbor (red), 
second-next-nearest-neighbor (green) and 
third-next-nearest-neighbor (blue) interactions. }
\label{fig0n}
\end{figure}       
                                
                             We implement    the   RSRG by iterating 
  the RG rules Eq. (\ref{jeff})  for each realization of  bare coupling parameters
   until the system has been completely decimated to one remaining effective bond whose energy excitation gap $\epsilon_1$  is recorded. The resulting distribution functions of such {\it exit gaps} for up to 300,000 random realizations are subsequently analyzed as a function of the decay exponent $\alpha$ and number of spins  N.                                 
                      
  In Fig. \ref{fig0n} we show   distribution functions 
                  of  $\epsilon_1$ for random XX chains ($J_z=0$, $N=128$) with antiferromagnetic  interactions  
                    of power $\alpha =0.5.$
                        Results for true long-range coupling (brown curve) are compared with short-ranged  couplings as obtained by 
                        including only 
 nearest-neighbor (black), 
next-nearest-neighbor (red), 
next-next-nearest-neighbor (green) and 
next-next-next-nearest-neighbor (blue)  in  Eq. (\ref{spinchain}).
\begin{figure}[t]
\includegraphics*[width=0.28 \textwidth,angle=-90]{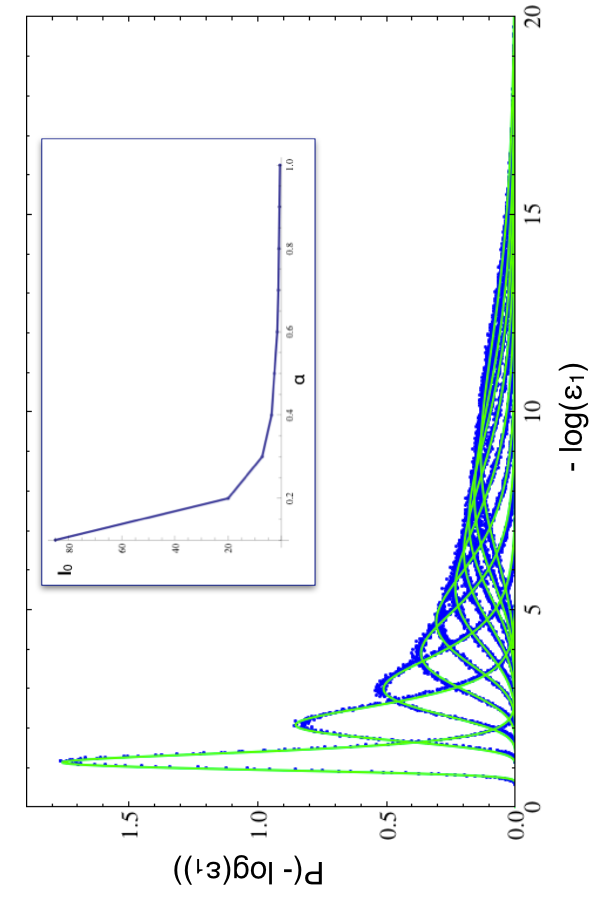} 
\vspace{0cm}
\caption{ {Real Space RG results for the distribution  
 of the lowest excitation energy gap $\epsilon_1$ 
  in the random XX-spin chain for $\alpha$  ranging from  to  $\alpha=0.1$ to  $\alpha=1.0,$
  where the interaction is cut off at 
nearest neighbors
   ($N=64$).  Green curve: IRFP distribution Eq. (\ref{weibull}) with $z =
                       \sqrt{N a_0/l_0}/\ln N,$ which indicates anomalous localization.  The fitted values of   $l_0$  in units of $a_0$ 
   are $l_0=85,20,7.2,3.7,2.6,1.5,1.1,.9,.7,.5$ from left to right,
as shown in the inset as function of $\alpha$.}}
\label{nearestneighbor}
\end{figure}          
   If   interactions   range only to  nearest-neighbors, 
        this model is known to have an 
            infinite-randomness fixed point (IRFP)\cite{refael}.         
    The distribution of  $\epsilon_1$  of random spin chains of N spins 
    with nearest-neighbor interactions  is then known to be {well } described  by the Weibull distribution\cite{juhasz}, 
                       \begin{equation} \label{weibull}
                         P_W(\epsilon_1) =\frac{u_0^{1/z} N}{z} \epsilon_1^{1/z-1}
                          \exp (- (u_0 \epsilon_1)^{1/z} N),
                       \end{equation}
                       where $0< z < \infty.$ 
                        $P_W$ is normalized,   $\int_0^{\infty} d \epsilon_1 P_W(\epsilon_1) =1$, and the expectation value 
                         of the excitation energy is 
 $            \langle \epsilon_1\rangle_W = \frac{\Gamma(1+z)}{u_0} N^{-z}$, where
  $\Gamma(x)$ is the Gamma-function. 
                  At the infinite randomness fixed point (IRFP) $z$ 
                   goes with system size $N$ to infinity. 
                     $z \rightarrow 
                       \sqrt{N a_0/l_0}/\ln N$  yields  for 
                     the expectation value of the Weibull function, 
                        $\langle \epsilon_1 \rangle \sim \exp (- \sqrt{ N a_0/l_0}),$
                       which {is known to be  the typical value of} the excitation energy of model 
                        Eq. (\ref{jw}) with random  nearest neighbor hopping{\cite{monthus,fisheryoung}}. 
 We checked that  nearest-neighbor results (black curves) are for all $\alpha >0$ well modeled by the  IRFP distribution 
   Eq. (\ref{weibull}) with $z =
                       \sqrt{N a_0/l_0}/\ln N$.
                       This is demonstrated 
     in Fig. 2  for values of $\alpha=0.1$  to $\alpha=1$ in steps of $0.1$ for $N=64$, where we find that   the length scale $l_0$ in units 
      of the initial distance between the spins $a_0$ increases continously when decreasing 
                       $\alpha$ as shown in the inset. This is due to the fact that 
                         the initial distribution of exchange couplings narrows with 
                         decreasing $\alpha,$ so that the flow to the strong disorder fixed point occurs at larger length scales $L>l_0.$
                       The parameter $u_0$ which shifts 
                        the center  of the distribution function, is fitted and found to decrease as $\alpha$ is increased. 
     
\begin{figure}[t]
\includegraphics*[width=0.45 \textwidth]{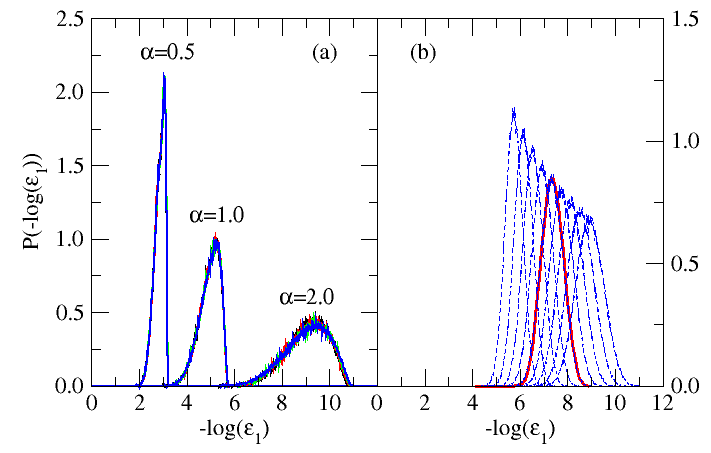} 
\caption{Distributions of
 $\epsilon_1$ 
  in  random long-range coupled XX-spin chains for (a) $\alpha=0.5,1.0,2.0$  and 
   $N=32, 64,128,256$  (black, red, green, blue),
     rescaled by their first moments.
(b)   $\alpha=1.2$  to $\alpha=2$ (from left to right) in steps of $.1$,
   $N=128$. Red Line:  critical  function  Eq. (\ref{critical})
                            multiplied  with a cutoff function,  $ \exp(-c/(x_{max}-x))
                      $, $c=4$ for $\alpha =1.6$. }
\label{figlr}
\end{figure}       

     {     The distribution changes strongly 
          when  couplings to farther sites are added\cite{zhou}.
          Chains with both nearest and next nearest neighbor couplings correspond
           to zig-zag chains, which have been studied
            in Ref. \cite{melin}, where it was concluded that the 
             distribution is given by Eq. (\ref{weibull}) with a finite $z$. 
              Similarly, one may expect that chains with further, but finite 
               range couplings are given by  Eq. (\ref{weibull}) with 
               another finite $z$. 
                We note  that $z=1$ corresponds to the distribution function of 
                 localized levels, the Poisson level spacing distribution $P(\epsilon_1)
                 = \exp (- \epsilon_1/\Delta)$. }
             {  It is well known that the level spacing of 
                    extended states is  well described  by the  Wigner surmise, which for time reversal invariant systems (GOE) gives
           \begin{equation} \label{wdo}
                P_{GOE}(\epsilon_1) =\frac{\pi}{2} \frac{\epsilon_1}{\Delta^2} \exp(-\frac{\pi}{4}(\epsilon_1/\Delta)^2).
                \end{equation}
                 $P_{GOE}$ is normalized such that 
                  $\int_0^{\infty} d \epsilon_1 P_{GOE}(\epsilon_1) =1$ and
                     $\int_0^{\infty} d \epsilon_1 \epsilon_1 P_{GOE}(\epsilon_1) =\Delta.$
                   At the delocalization transition   critical states  are known to obey 
                critical level spacing distributions,  
                 conjectured to be of the form}\cite{cuevas},  
                 \begin{equation} \label{critical}
                 P_c(\epsilon_1/\Delta) = B (\epsilon_1/\Delta^2)  \exp (- A (\epsilon_1/\Delta)
               ^{\beta}  ).
               \end{equation}
               Note that 
                 this distribution is normalized 
                   $\int_0^{\infty} d x P_c (x)=1$
 and the first moment  is $\langle \epsilon_1 \rangle = \Delta.$
  { We will further analyse  random spin chains with a finite range interaction in Ref. \cite{moure}.
                 Here, we 
                 concentrate on  random spin chains with  truly  long-range, antiferromagnetic 
                   interactions. }
                 In Fig. \ref{figlr} we show  results  for  distribution functions 
                  of $\epsilon_1$, scaled by their first moments.  
                  We observe that by lowering the decay exponent $\alpha$,  
                  rendering the interactions more long-ranged, the distribution functions are moved towards higher energies and  become more narrow.
                   Strikingly,  a sharp cutoff  is observed at large 
                    $x= -\log(\epsilon_1)$ for all   $\alpha$ considered. This cutoff  coincides exactly with 
                      the excitation energy of a singlet of spins which are  coupled by the  bare coupling at  maximal distance $L/2$,
                      which for  XX chains depends on chain length $L$ as 
                      $\epsilon_{min} = 1/2 J (L/2)^{-\alpha}$. Thus, that  lower limit 
                        to the excitation energy  has to be taken into account  in deriving the distribution function.
               {    Therefore, we conjecture that for all $\alpha$ the distributions are modifed by  a} cutoff factor 
                   of  form $ \exp(-c/(x_{max}-x)),
                      $
                          with an essential singularity at $x_{max}=\alpha
                         {\rm log} (L/2)) + {\rm log} (2/J)$.

                 \begin{figure}[t]
\includegraphics*[width=0.48 \textwidth]{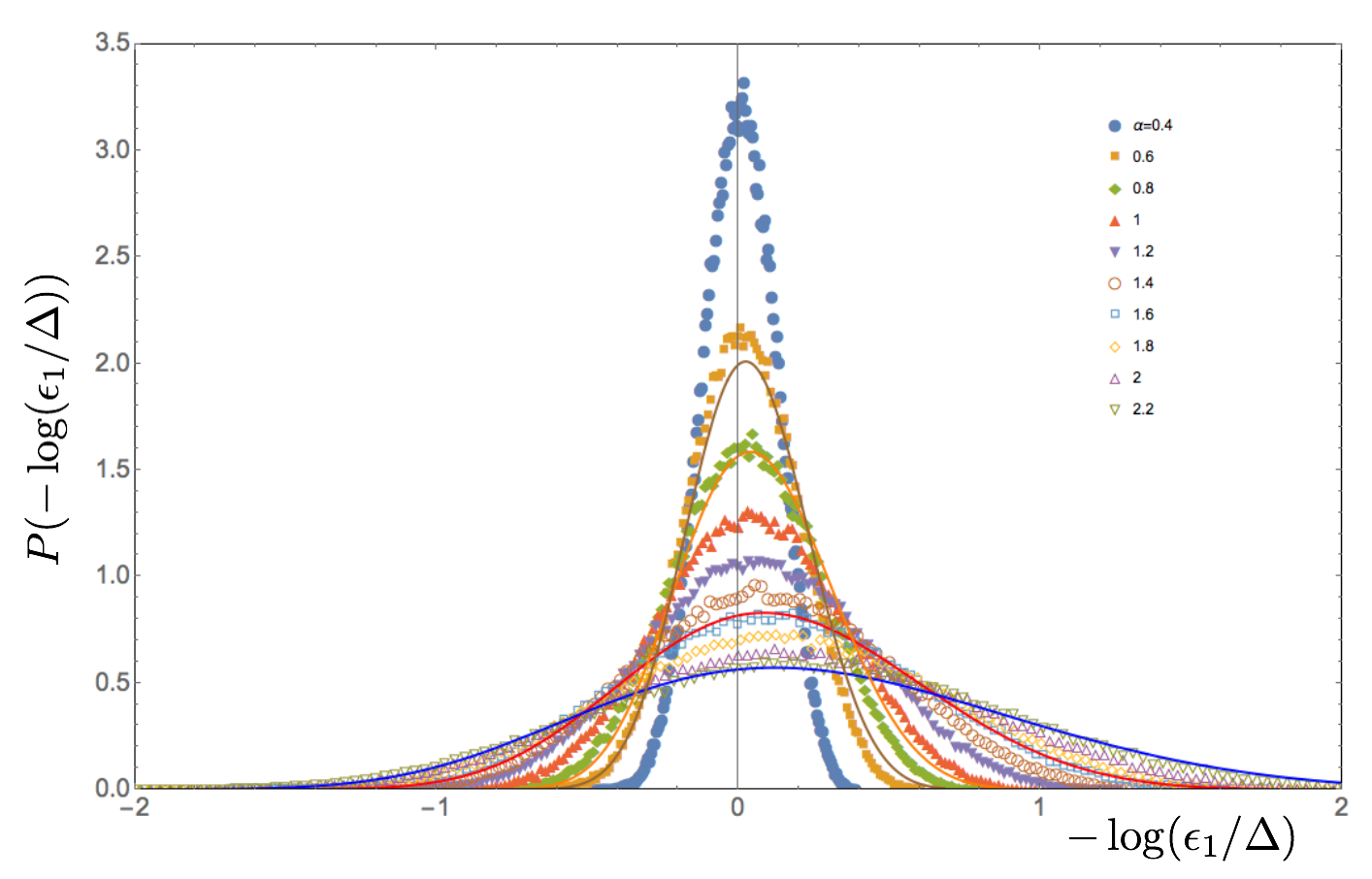} 
\caption{{ Distribution function for N=320 
 together with the critical distribution function   multiplied  with the cutoff function (c=8) with $x_{max}(L=3200)$  for $\alpha= 1.6$ and normalized (red curve), 
                       the Poisson level distribution function  (blue curve)
                                       for $\alpha =2.2$, the orthogonal Wigner surmise 
                                         for  $\alpha= .8$     (orange curve), and 
                                       the unitary  Wigner surmise    for  $\alpha= .6$ (brown curve),
                                     all    
                                      multiplied with the same cutoff function (c=8) and normalized.  }
}
\label{figlrt}
\end{figure}

                           In Fig. \ref{figlr}  (b) we show  numerical results 
                            for $\epsilon_1$  as obtained from the RSRG from the 3rd last RG step\cite{remark}
                             together with a plot of 
                            the analytical critical distribution functions  Eq. (\ref{critical}) with $ \Delta = N^{-\alpha}$,   {with  $ \beta =1$, $A=2$ and $B=4$ (red curve in Fig.  \ref{figlr} b)),
                            multiplied  with the cutoff function (c=4) and normalized.}
                         {   We find that for $N=128$ }  it   fits the data very well at $ \alpha_c=1.6$ (red).
                               Thus, this crossover in the 
                                level distribution function  {could be an } indication of   a many-body localization-delocalization transition at $ \alpha_c=1.6$.
                               {  We performed
                                  the calculations also for larger 
                                   system sizes $N=320.$ The result is 
                               shown in Fig.  \ref{figlrt} together with  the 
                                     critical distribution function     multiplied  with the cutoff function (c=8) with $x_{max}(L=3200)$  and  normalized (red curve) fits the data well for $\alpha= 1.6$. 
                             As additional evidence
                                for the delocalization transition we plot 
                                        in Fig.  \ref{figlrt}                              the Poisson level distribution function 
                                      multiplied with the same cutoff function (c=8) and normalized (blue curve)
                                       for $\alpha =2.2$, which confirms that states are localized there, while 
                                         for 
                                        $\alpha= .8$ the data fits  the orthogonal Wigner surmise multiplied with the cutoff function 
                                           (orange curve), indicating that the states are extended. 
                                         For $\alpha= .6$  the unitary  Wigner surmise (brown curve)
                                        multiplied with the cutoff function  fits the data well.  
                                        This can be explained by the fact that  the  phase factors in Eq. (\ref{jw}),  correspond to  random magnetic field factors, when neglecting their quantum dynamics, 
         which    break  time reversal symmetry and change the universality class  to the unitary one. 
                                          For $\alpha < 1$  the density of states at small energies becomes more sparse
                                          and deviations from the Wigner surmise are seen. 
                                          This can be an indication of a reduction of the density of states,
                                            due to the   long range interactions.
                                         In fact a mean field theory analysis   
                                         yields a logarithmic Coulomb gap for $\alpha = d=1$
                                            and a power law pseudogap for $\alpha < 1$\cite{gap,Yu}. }

                                Turning on the z-coupling towards the isotropic  Heisenberg chain the 
          RG rules are still given by Eq. (\ref{jeff}) and we do not find a qualitative 
          change of this delocalization physics. The excitation gap becomes enhanced
           by a factor $2$  changing only  the quantitative  value of $\alpha_c$. 
      In the  Ising limit, the renormalization rules  change \cite{isingrgrules} since the ground state becomes degenerate.
       Random Ising spin chains with long-range coupling have  been studied  in Ref. \cite{katzgraber},
        where it was found that all excitations are short-ranged for $\alpha>2$,
         while  a correlated  spin glass phase with $T_c=0$ exists for  $1<\alpha<2$.
    { Recently, a quantum Ising model with long range interactions
      has been considered
      finding a strong disorder fixed point with a dynamical 
       exponent $z= \alpha$  \cite{kovacs}.}
            
{\bf Conclusions and Discussion. --}
 In conclusion, we   analyzed the full  distribution functions  of the first excitation energy 
  $\epsilon_1$ from the ground state
  of  the anisotropic Heisenberg model
with long-range interactions, decaying with a power $\alpha$, as generated by placing spin sites at random 
positions along the chain. 
  In the XX-regime we find  
   critical values   $\alpha_c$ where  the  distribution of excitation energies coincides with the critical distribution  function, 
 { indicating  that for smaller  $\alpha< \alpha_c$  there is an  extended  phase}.  

    We will    in Ref. \cite{moure}
    perform a {more } extensive finite size scaling analysis of the spectral statistics 
    using methods developed previously\cite{levelstatistics} in order to  establish  the 
     criticality and to extract the  critical parameters.

  Since one of our motivations for this work
    was to contribute to an understanding of the
     anomalies in the   magnetic properties of  doped 
 semiconductors, we may  ask if 
 the  crossover in the distribution function of excitation energies 
           can be experimentally  detected by measuring the temperature dependence
            of the magnetic susceptibility
              $\chi (T) = n_{FM}(T)/T$,
             where $ n_{FM}(T)$ is  the number of free magnetic moments
              at temperature $T$.
              At the IRFP, one obtains with the singular density of states,
           $\rho(\epsilon) = |\epsilon|^{-1} \ln |\epsilon|^{-3}$, that $n_{FM}(T) = n_M \int_0^T 
           d \epsilon \rho (\epsilon) = n_M/\ln^2 T$.
            Thus, one finds  a logarithmic reduction of the paramagnetic 
            susceptibility.  We expect 
             when interactions range beyond nearest neighbor  a
             crossover to Poissonian distribution 
             and to the
             constant density of states   $\rho(\epsilon) = \rho_0$. Accordingly, 
              one expects that  $\chi(T) \rightarrow n_M \rho_0$, 
               which is    independent 
               of temperature. Thus, 
                we conclude that the 
                suppression of the singularity of the density of states, 
                 when increasing the range of the interaction, 
                  weakens the singularity in the magnetic susceptibility. Thus, 
                   we may expect a  decrease  of 
                    the magnetic susceptibility as the doping is increased  from the insulator towards
                    the metal phase. 
               On the other hand,  the many body delocalization transition in the spin degrees of freedom 
                cannot be observed in  such thermodynamic measurements 
                   since the average density of states
                is unchanged at  this quantum phase transition. Rather, in order to detect the delocalization transition  we suggest 
                  time dependent measurements such as spin echo experiments.

         \acknowledgments

We would like to acknowledge Cecile Monthus  for useful discussions and Bruce Normand for critical reading and useful comments. The numerical computations were carried out on the University of Southern California High Performance Supercomputer Cluster. This research has been supported by the Department of Energy Grant No. DE-FG02-05ER46240 and the DFG KE-15 Collaboration grant. S. Haas would also like to thank the Humboldt Foundation for support.


%

\begin{thebibliography}*


\bibitem{pwanderson0} P. W. Anderson, Phys. Rev. {\bf 109}, 1492 (1958).

\bibitem{pwanderson}  P. W. Anderson, Nobel Lectures in Physics {\bf 1980}, 376
(1977).

\bibitem{mott} N. F. Mott, J. Phys. Colloques 37, C4 (1976).

\bibitem{loehneysen}
 H. v. L\"ohneysen, Adv. in Solid State Phys. {\bf 40}, 143 (2000).
 
 \bibitem{finkelsteinesr} A. M. Finkel'shtein, JETP Lett. {\bf 46},
 513 (1987). 
 
 \bibitem{senthil} A. C. Potter, M. Barkeshli,  J. McGreevy,  T.  Senthil, Phys. Rev. Lett. 
{\bf 109}, 077205
(2012). 


\bibitem{rkky} M. A. Ruderman and C. Kittel, Phys. Rev. {\bf 96}, 99 (1954), T. Kasuya, Progress of Theoretical Physics {\bf 16}, 45 (1956); K. Yosida, Phys. Rev. {\bf 106}, 893 (1957).

   
   \bibitem{reviewtls}D. Salvino, S. Rogge, B. Tigner, D. Osheroff,
Phys.Rev.Lett. {\bf  73}, 286 (1994).

\bibitem{Yu} C. C. Yu, A. J. Leggett, Commun. Condens. Mat. Phys.
{\bf 14}, 231 (1988).


\bibitem{levitov} L. S. Levitov, Ann. Phys. {\bf 8}, 507 (1999).

 \bibitem{prl1999} 
   S. Kettemann, P. Fulde, P. Strehlow, Phys. Rev. Lett. {\bf 83}, 4325 (1999).	


\bibitem{dasguptama} C. Dasgupta, S.-K. Ma, Phys. Rev. B {\bf 22}, 1305 (1980). 


\bibitem{bhattlee} R. N. Bhatt,  P. A. Lee, Phys. Rev. Lett. {\bf 48}, 344 (1982). 



\bibitem{fisher} D. S. Fisher, Phys. Rev. Lett. {\bf 69}, 534 (1992).

\bibitem{fisher2} D. S. Fisher, Phys. Rev. B {\bf 50}, 3799  (1994).


\bibitem{bhattfisher} R. N. Bhatt and D. S. Fisher, Phys. Rev. Lett. {\bf 68}, 3072 (1992).


\bibitem{levelstatistics}
B. L. Altshuler and B. I. Shklovskii, Sov. Phys. JETP 64, 127 (1986);
U. Sevan, Y. Imry, Phys. Rev. B 35, 6074 (1987);
B. L. Altshuler, I. Kh. Zharekeshev, S. A. Kotochigova, and B. I. Shklovskii, Sov.
Phys. JETP 67, 625 (1988); I. Kh. Zharekeshev, Sov.Phys. Solid State 31, 65 (1989);
 S. N. Evangelou and E. N. Economou, Phys. Rev. Lett. 68, 361 (1992);
 B. I. Shklovskii, B. Shapiro, B. R. Sears, P. Lambrianides, and H. B. Shore, Phys.
Rev. B 47, 11487 (1993);
 E. Hofstetter and M. Schreiber, Phys. Rev. B 48, 16979 (1993); 49, 14726 (1994);
 	I. Kh. Zharekeshev, B. Kramer, Phys. Rev. B 51, 17239 (1995).




\bibitem{fyodorov1996}  A. D. Mirlin, Y. V. Fyodorov, F.-M. Dittes, J. Quezada, T. H. Seligman, Phys. Rev. E 54, 3221 (1996).

\bibitem{efetov} K. B. Efetov, Supersymmetry in Disorder and Chaos, Cambridge
University Press, Cambridge, 1997.

\bibitem{mackinnon} B. Kramer and A. MacKinnon, Rep. Prog. Phys. {\bf 56},
1469 (1993).

\bibitem{rmpmirlinevers} F. Evers and A. D. Mirlin,
{\it Rev. Mod. Phys.} {\bf 80}, 1355 (2008).


\bibitem{multifractal} F.Wegner, Z. Phys. B 36, 209 (1980); H. Aoki, J. Phys. C 16, L205
(1983); C. Castellani and L. Peliti, J. Phys. A 19, L991 (1986);
M. Schreiber and H. Gru\ss bach, Phys. Rev. Lett. 67, 607 (1991);
M. Janssen, Int. J. Mod. Phys. B 8, 943 (1994).

\bibitem{nonrandomlr}
A. Rodríguez, V. A. Malyshev, G. Sierra, M. A. Martín-Delgado, J. Rodríguez-Laguna, and F. Domínguez-Adame
Phys. Rev. Lett. 90, 027404  (2003);  A. V. Malyshev, V. A. Malyshev, and F. Domínguez-Adame
Phys. Rev. B 70, 172202 (2004).

\bibitem{hierarchical}
  F. L. Metz, L. Leuzzi, G. Parisi, and V. Sacksteder, IV
Phys. Rev. B 88, 045103 (2013).



	\bibitem{monthus}
F. Igloi, C. Monthus, 
Phys. Rep.  {\bf 412}, 277 (2005). 

\bibitem{bruno}  P. Bruno, Phys. Rev. Lett. {\bf 87}, 137203 (2001).



  \bibitem{dyson} F. J. Dyson, 
Phys. Rev. {\bf 92}, 1331(1953).

\bibitem{wegner} F. J. Wegner, in {\it Fifty Years of Anderson Localization},
World Scientific (2010);  arXiv:1003.0787.



  \bibitem{refael} G. Refael, E. Altman, Comptes Rendus Physique, Vol. 14, Issue 8, 725-739 (2013).
  
 

\bibitem{damle} O. Motrunich, K. Damle, D. A. Huse, Phys. Rev. B {\bf 63}, 224204 (2001).



\bibitem{haas1}
S. Haas, J. Riera, and E. Dagotto, Phys. Rev. B {\bf 48}, 13174 (1993) (RC).



 

\bibitem{sigrist} E. Westerberg, A. Furusaki, M. Sigrist, and P. A. Lee,  Phys. Rev. B {\bf 55}, 12578 (1997). 



\bibitem{isingrgrules}   $
               J_{z lm}^{eff} =   J_{z lm} - \frac{(J_{z li}-J_{z lj})(J_{z im}-J_{z jm})}{\sqrt{(J_{z li}-J_{z lj})^2 + (J_{z im}-J_{z jm})^2  +4 J_{x ij}^2   }},
              $
              while $  J_{x lm}^{eff}$ is still given by Eq. (\ref{jeff}).
  
  

\bibitem{basko} D. M.  Basko,  I. L.  Aleiner, B. L.  Altshuler, Annals of Physics,  {\bf 321}, 1126 (2006). 

\bibitem{znidaric} M. Znidaric, T. Prosen, P. Prelovsek,Phys. Rev. B 77, 064426 (2008).

\bibitem{garel} C. Monthus, T. Garel, Phys. Rev. {\bf  B 81},134202 (2010). 

\bibitem{bardarson} J. H. Bardarson, F. Pollmann, J. E. Moore,
Phys. Rev. Lett. {\bf 109}, 017202 (2012).
  



\bibitem{juhasz} R. Juhasz and Y.-C. Lin,
F. Igloi, Phys. Rev. B 73, 224206 (2006).
{
\bibitem{fisheryoung} D.S. Fisher, A.P. Young, Phys. Rev. {\bf B 58}, 9131 (1998).
 In this reference the authors find a different scaling for the 
  expectation value of the excitation energy in the random short range quantum Ising model, 
    $\langle \epsilon_1 \rangle \sim \exp (-c N^{1/3}).$ }

\bibitem{zhou}  C.G. Zhou and R. N. Bhatt, 
Phys. Rev. B 68, 045101 (2003).
{
\bibitem{melin}  
R. M\'elin, Y.-C. Lin, P. Lajk\'o,3 H. Rieger, and F. Iglo\'i, Phys. Rev. 
B {\bf 65}, 104415 (2002).}



\bibitem{cuevas}
E. Cuevas, 
Europhys. Lett. 67, 84 (2004)


\bibitem{moure} N. Moure, S. Kettemann, S. Haas, unpublished (2015). 




\bibitem{remark}  Here, we took the excitation energy of the strongest singlet pair of 
 the 3rd last RG step. Accordingly $x_{\rm max} = \log[2] + \alpha (\log L/6)$.
  We found that  the distribution of the excitation energy  from the singlet in the last RG step of the 
    chain with true long-ranged coupling  is masked
    too much by the cutoff function, so that one cannot determine $\alpha_c$ accurately.
    {
\bibitem{gap} A. L. Efros and B. I.  Shklovskii,
J. Phys. {\bf C 8}, L49 (1975).}


\bibitem{katzgraber} H. G. Katzgraber,  J. Phys. Conf. Ser. {\bf 95}, 012004 (2008).


 	{
\bibitem{kovacs} R. Juh\'asz, I. A. Kov\'acs and F. Igl\'oi,
 Europhys. Lett. {\bf 107}, 47008 (2014); C. Monthus, 
J. of Stat. Mech.   P05026  (2015). }





%

\end{thebibliography}
\end{document}